\documentclass[aps,pre,groupedaddress,twocolumn,showpacs,floatfix]{revtex4}
\usepackage{epsfig} 

\begin{document}

\title{
Statistics of active vs. passive advections in magnetohydrodynamic
turbulence}   
\author{Thomas Gilbert}
\email[]{thomas.gilbert@inln.cnrs.fr}
\affiliation{Institut Non-Lin\'eaire de Nice, CNRS, Universit\'e de Nice,
1361 Route des Lucioles, 06560 Valbonne, France}
\author{Dhrubaditya Mitra}
\email[]{dhruba@physics.iisc.ernet.in}
\affiliation{Centre for Condensed Matter Theory, Department of Physics,
              Indian Institute of Science, Bangalore 560 012, India}
\date{\today}

\begin{abstract}
Active turbulent advection is considered in the context of
magneto-hydrodynamics. In this case, an auxiliary passive field bears no
apparent connection to the active field. The scaling  
properties of the two fields are different. In the framework of a shell
model, we show that the two-point structure function of the passive field
has a unique zero mode, characterizing 
the scaling of this field only. In other words, the existence of
statistical invariants for the decaying passive field carries no information
on the scaling properties of the active field.
\end{abstract}
\maketitle

In the context of turbulent advection, the understanding of fluid turbulence 
has greatly improved in the recent years \cite{fgv01}. The anomalous scaling 
has been shown to be universal and connected to the existence of statistical
integrals of motion \cite{cv01}. In \cite{abcpv01}, it was shown that the 
statistically conserved structures of decaying passive turbulence 
dominate the statistics of forced turbulence, thus offering a rather general
framework for understanding the universality of anomalous scaling in forced
turbulence.

Let $\phi$ be a decaying field transported by a stationary turbulent flow. The
linearity of the advection implies the following relation for the correlation
functions~:
\begin{eqnarray}
\label{propcorf}
\lefteqn{\langle \phi(\vec{r}_1, t) \dots \phi(\vec{r}_N, t)\rangle =}&&\nonumber\\
&&\int d\underline{\vec{q}} \mathcal{P}^{(N)}(\underline{\vec{r}},t 
|\underline{\vec{q}},t_0) 
\langle \phi(\vec{q}_1, t_0) \dots \phi(\vec{q}_N, t_0)\rangle \ ,
\label{tevolcf}
\end{eqnarray}
where we used the compact notation $\underline{\vec{r}}\equiv
\vec{r}_1,\dots,\vec{r}_N$ to denote a collection of $N$ position vectors.
Equation (\ref{tevolcf}) tells us there exists a linear operator 
$\mathcal{P}^{(N)}$ that propagates the 
$n$th order correlation function from time $t_0$ to time $t$. Without fresh 
input, that is in the absence of forcing, the correlation functions of $\phi$
decay due to dissipative effects. Nevertheless, as conjectured in 
\cite{abcpv01}, there exist special functions $Z^{(N)}$ that are left 
eigenfunctions of eigenvalue $1$ of the operator $\mathcal{P}^{(N)}$,
\begin{equation}
Z^{(N)}(\underline{\vec{r}}) = \int
d\underline{\vec{q}}\mathcal{P}^{(N)}(\underline{\vec{q}},t 
|\underline{\vec{r}},t_0) Z^{(N)}(\underline{\vec{q}}),
\label{leftev}
\end{equation}
such that
\begin{equation}
I^{(N)}(t) = \int d\underline{\vec{r}}
Z^{(N)}(\underline{\vec{r}}) \langle \phi(\vec{q}_1,t) \dots 
\phi(\vec{q}_N,t)\rangle
\label{prestr}
\end{equation}
is preserved in time. $I^{(N)}$  and $Z^{(N)}$ are respectively called 
a statistical integral of motion and a statistically
preserved structure of order $N$, also referred to as zero modes
\footnote{Note that Eq. (\ref{leftev}) is not stricto sensu an eigenvalue
  problem; $\mathcal{P}^{(N)}$ is not defined on a compact space. Only in the
  case of a scaling advecting field, is it possible to write down a proper
  operator and identify the zero modes as left eigenmodes with unit
  eigenvalues \cite{bgk98}.}. 

Now, consider the same passive advection problem with an external forcing,
such that the system reaches a
stationary state. Define,
the correlation function of $\phi$ in that stationary state to be,
\begin{equation}
F^{(N)}(\underline{\vec{r}}) = 
\langle \phi(\vec{r}_1, t) \dots \phi(\vec{r}_N, t)\rangle_f,
\end{equation}
 where the symbol $ \langle \cdot \rangle_f$ denotes averaging over the
statistical stationary state.   
It was conjectured in Ref~\cite{abcpv01}, that the anomalous part of 
$F^{(N)}(\underline{\vec{r}})$  is dominated by the leading 
zero modes of the decaying problem, 
i.~e. $Z^{(N)} \sim F^{(N)}$. 
The conjecture was
verified in the context of a shell model for passive scalar advection.

In subsequent studies, it was discovered that the existence of statistical 
invariants of the motion for passive turbulence may help understand the 
statistics of active turbulence, a case where the advected quantity affects 
the dynamics of the advecting field. 
In \cite{cmmv02} and \cite{ccgp02,ccgp03}, the case of 
thermal convection in the Boussinesq approximation was studied. There it
was shown 
that the scaling of the active field is also dominated by the statistically 
preserved structures of auxiliary passive fields. It is yet unclear how 
general this connection between the statistics of active and auxiliary 
passive fields is. The case of 2-dimensional magneto-hydrodynamics is 
revealing. Indeed, in this case active and passive fields have very different 
scaling behaviors. As shown in \cite{ccmv02}, the transported fields
cascade in different directions. It is therefore unexpected  
that the statistics of the auxiliary passive field holds information on the 
statistics of the active field. The two fields have different scaling 
properties.  In \cite{ccgp03}, despite this difference, the claim was made 
that the analogy does hold in the sense that there exist sub-leading 
zero modes of the propagator of the correlation functions of the auxiliary 
passive field with the scaling of the correlation functions of the active 
field.

The purpose of this note is to show that this is actually not the case. To
this end, we will limit our investigation to the case of the second
order structure  
functions of a shell model of 2-dimensional magneto-hydrodynamic turbulence. 
The case of the second order structure function is the simplest
one. Because of the absence of geometry, its scaling is
non-anomalous \cite{cv01,gz98,bgk98}, which in the language of zero
modes implies non-degeneracy, that is there is a unique conserved structure
associated to the two-point statistical invariant. The same holds in the
language of shell models, where the only two point function of a passive
scalar field $\theta_n$ associated to the scale $k_n$ is
$\langle|\theta_n|^2\rangle$, in contrast to higher order structure
functions, e.~g. the fourth order for which we have contributions from 
$\langle|\theta_n|^4\rangle$,
$\langle|\theta_{n-1}|^2|\theta_{n+1}|^2\rangle$, etc. 
Using methods similar to those used in \cite{cgp02}, we will construct the 
operator propagating the second order structure functions, and will demonstrate
that the auxiliary passive field has the same statistical integral of motion 
as other shell models of passive advection. It will be inferred that there
is no sub-leading zero-mode with the scaling of the (active) magnetic field.

In analogy to other models \cite{basu98,frick98,g99}, the following two
sets of equations  
generalize the usual Sabra shell model \cite{lpppv98} for the turbulent 
velocity field to magneto-hydrodynamic turbulence
(we omit dissipative terms)~:
\begin{eqnarray}
\frac{d u_n}{dt} &=& i [k_{n+1} (u_{n+1}^*u_{n+2}-b_{n+1}^*b_{n+2})
\nonumber\\&&
- (\epsilon +1) k_n (u_{n-1}^*u_{n+1} - b_{n-1}^*b_{n+1})\label{mhdsabrau}
\\
&&- \epsilon k_{n-1}(u_{n-2}u_{n-1} - b_{n-2}b_{n-1})] + f_n ,\nonumber
\\
\frac{d b_n}{dt}&=& i [-(\epsilon+\delta) k_{n+1}
(u_{n+1}^*b_{n+2}-b_{n+1}^*u_{n+2}) \nonumber\\&&
+ \delta k_n (u_{n-1}^*b_{n+1} - b_{n-1}^*u_{n+1})\label{mhdsabrab}
\\
&&+ (\delta -1) k_{n-1}(u_{n-2}b_{n-1} - b_{n-2}u_{n-1})] + f'_n .
\nonumber
\end{eqnarray}
Here $u_n$ (the velocity field) and $b_n$ (the magnetic field) are complex
variables defined on a discrete set of shells indexed by the integer $n$ whose
associated wave-number $k_n = k_0 \lambda^n$, $\lambda>1$ (hereafter taken to 
be  $\lambda=2$). $f_n$ and $f'_n$ are two forcing terms which are taken to be
stochastic white noises with identical statistics and concentrated on a limited
number of neighboring shells ($n=5,6,7$ in our numerical experiments). The
model's parameters $\epsilon$ and $\delta$ are conveniently parametrized 
in the following way. The three-dimensional model for which $\epsilon < 0$ 
reads ($\alpha>0$)
\begin{equation}
\begin{array}{l}
\epsilon = - \lambda^{-\alpha}\ ,\\
\delta = (1 + \lambda^{\alpha})^{-1}\ .
\end{array}
\quad
\mbox{(3D model)}
\label{3dparam}
\end{equation}
The two-dimensional problem on the other hand has $\epsilon>0$ and reads
\begin{equation}
\begin{array}{l}
\epsilon = \lambda^{-\alpha}\ ,\\
\delta = -(\lambda^{\alpha}-1)^{-1}\ .
\end{array}
\quad
\mbox{(2D model)}
\label{2dparam}
\end{equation}
Correspondingly, we have the following quadratic dynamical invariants
(i.~e. time-invariant in the limit of zero viscosity and zero external
forcing).  
\begin{eqnarray}
E &=& \sum_n (|u_n|^2 + |b_n|^2)\quad\mbox{(total energy)}, \label{toten}\\
K &=& \sum_n \Re(u_n^* b_n)\quad\mbox{(cross helicity)},\label{crosshe}\\
H &=& \sum_n \mathrm{sign}(\delta)^n k_n^{-\alpha} |b_n|^2 
\quad\mbox{(magnetic helicity)}.
\label{magnhe}
\end{eqnarray}
Thus, in both two- and three-dimensional models, the equations 
(\ref{mhdsabrau}, \ref{mhdsabrab}) have one single free parameter, 
$\alpha>0$.  The two- and three-dimensional models actually have very 
different dynamical behaviors, see \cite{g99}. 
Only the two-dimensional model can sustain a stationary state and we will
limit ourselves to this case.
As dimensional analysis shows \cite{ccgp03}, the conservation of the first two
invariants implies that $u_n$ and $b_n$ must both have Kolmogorov scalings,
$\langle |u_n|^2\rangle, \langle |b_n|^2\rangle \sim k_n^{-2/3}$,
for which the corresponding fluxes are constant. This 
is indeed what has been measured for similar models \cite{g99}, where both 
fields appear to display the same anomalies. Further, as shown in 
\cite{ccgp03}, the conservation of the third invariant allows for another 
scaling, $\langle |b_n|^2\rangle \sim k_n^{\alpha-2/3}$, for which the 
magnetic helicity flux is constant. But since this scaling is incompatible 
with the conservation of the two other invariants, it is not relevant to the 
statistics of the magnetic field.

However if one considers a passive auxiliary field obeying an equation 
identical to Eq. (\ref{mhdsabrab}) for the 2-dimensional case,
\begin{eqnarray}
\frac{d a_n}{dt}&=& i [\lambda^{-\alpha}(\lambda^\alpha-1)^{-1} k_{n+1}
(u_{n+1}^*a_{n+2}-a_{n+1}^*u_{n+2}) \nonumber\\
&&- (\lambda^\alpha-1)^{-1} k_n (u_{n-1}^*a_{n+1} - a_{n-1}^*u_{n+1})\nonumber\\
&& - \lambda^\alpha(\lambda^\alpha-1)^{-1} k_{n-1}(u_{n-2}a_{n-1} 
- a_{n-2}u_{n-1})]\ ,\nonumber\\
\label{mhdsabraa}
\end{eqnarray}
the only relevant invariant is the equivalent of the magnetic helicity 
Eq. (\ref{magnhe}), to which a constant flux is associated. Thus the
dimensional scaling $\langle|a_n|^2\rangle \sim k_n^{\alpha-2/3}$
is expected to be observed. And that is indeed what was found in 
\cite{ccgp03}. 

Notice though that the linearity of Eq. (\ref{mhdsabraa}) allows for the
substitution $\psi_n = \lambda^{-\alpha n/2} a_n$. 
With this new variable, Eq. (\ref{mhdsabraa}) takes the form,
\begin{eqnarray}
\frac{d\psi_n}{dt} &=& i[A(k_{+1} u^*_{n+1}\psi_{n+2} +
k_{n-1}\psi_{n-2}u_{n-1}) \nonumber\\
&&+ B(k_{n+1}\psi^*_{n+1}u_{n+2} - 
k_n \psi^*_{n-1}u_{n+1}) \nonumber\\
&&+ C(k_n u^*_{n-1}\psi_{n+1} +
k_{n-1}u_{n-2}\psi_{n-1})],
\label{passcalar}
\end{eqnarray}
which describes the advection of a scalar for which the
quadratic invariant is $\sum_n |\psi_n|^2$, which is similar to the 
shell model with only nearest neighbor interaction considered in 
\cite{abcpv01,cgp02}.  The coefficients
in Eq. (\ref{mhdsabraa}) correspond to the choice
$A = (\lambda^\alpha-1)^{-1}$,
$B = - \lambda^{-\alpha/2}(\lambda^\alpha-1)^{-1}$, and
$C = - \lambda^{\alpha/2}(\lambda^\alpha-1)^{-1}$.
This shows that as far as shell models are concerned, the difference
between a passively advected vector and passively advected scalar
is just a numerical factor. Henceforth
we do not make a distinction between the two and use the name "passive
field" for both.   

\begin{figure}
\centerline{\psfig{figure=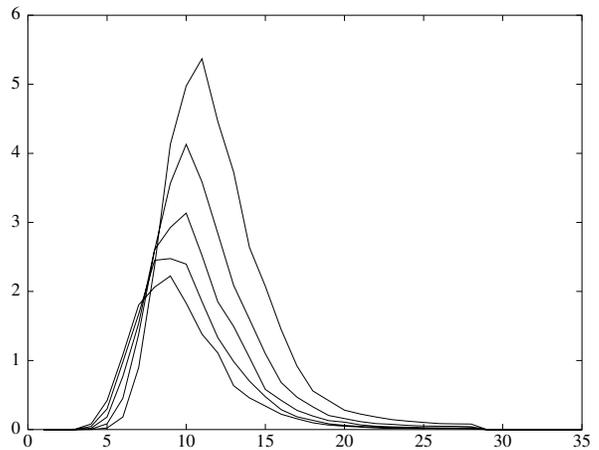,width=.7 \hsize,angle=-90}}
\caption{The elements of $\mathcal{P}^{(2)}_{n,20}$ for the scalar field
Eq. (\ref{passcalar}) advected by the MHD field Eqs. (\ref{mhdsabrau}, 
\ref{mhdsabrab}), where the parameter was chosen to be $\alpha=2$. 
The times displayed are $.4$, $.55$, $.7$, $.85$ and $1$ respectively
(measured in the natural time units of the model). The horizontal axis
corresponds to the shell numbers. The units on the vertical scale are
arbitrary. 
The simulation was done using a total of $35$ shells, with
the first shell wave-number $k_0 = 1/16$ and a shell spacing of $\lambda=2$.
All the fields were dissipated on the small scales with a term $\nu k_n^2$,
with $\nu = 10^{-12}$. The advecting fields were forced on shells $5-7$ with 
white delta correlated noise of amplitudes $1/\sqrt{2}$, $1/2\sqrt{2}$ and
$1/4$ respectively. Moreover the phase of the forcing on shell $7$ was taken 
to be equal to the sum of the phases of the forcings on shells $5$ and $6$.}
\label{fig.P2}
\end{figure}

Consider now the equivalent of Eq. (\ref{propcorf}) for the propagation 
of the second order structure functions $\langle |\theta_n|^2 \rangle$ in the 
decaying problem -- including dissipative terms on the RHS of Eq. 
(\ref{passcalar}).
Following notations similar to those used in \cite{cgp02}, we can write the 
equation of motion for $\psi_n$ under the form
\begin{equation}
\frac{d \psi_n}{dt} = \mathcal{L}_{n,m} \psi_m\ ,
\end{equation}
with the solution 
\begin{eqnarray}
\psi_n(t) &=& \mathsf{T}^+ 
\left\{\exp\left[\int_{t_0}^t ds \mathcal{L}(s)\right]\right\}_{n,m}
\psi_m(t_0), \nonumber\\
&\equiv& \mathcal{R}_{n,m}(t|t_0) \psi_m(t_0),
\end{eqnarray}
($\mathsf{T}^+$ denotes the time ordering operator). Letting
\begin{equation}
\mathcal{P}^{(2)}_{n,m}(t|t_0) \equiv \langle 
\mathcal{R}_{n,m}(t|t_0)\mathcal{R}^*_{n,m}(t|t_0) \rangle\ ,
\end{equation}
the propagation of second order structure functions obeys the following 
equation~:
\begin{equation}
\langle|\psi_n(t)|^2\rangle = \sum_m \mathcal{P}^{(2)}_{n,m}(t|t_0)
\langle|\psi_m(t_0)|^2\rangle\ .
\label{propa2}
\end{equation}
The form of the operator $\mathcal{P}^{(2)}$ was discussed in \cite{cgp02}. 
It is a matrix whose elements can be obtained by propagating an initial 
condition concentrated at a given shell. Similar considerations hold for the 
models considered here. In Fig. \ref{fig.P2} we show these elements for 
successive times, starting from an initial conditions at shell $20$. The model
we used is Eq. (\ref{passcalar}) for the parameters corresponding to Eq. 
(\ref{mhdsabraa}) and advected by the magneto-hydrodynamic fields Eqs. 
(\ref{mhdsabrau}, \ref{mhdsabrab}). The parameters of the simulation are given 
in the figure caption.

\begin{figure}
\centerline{\psfig{figure=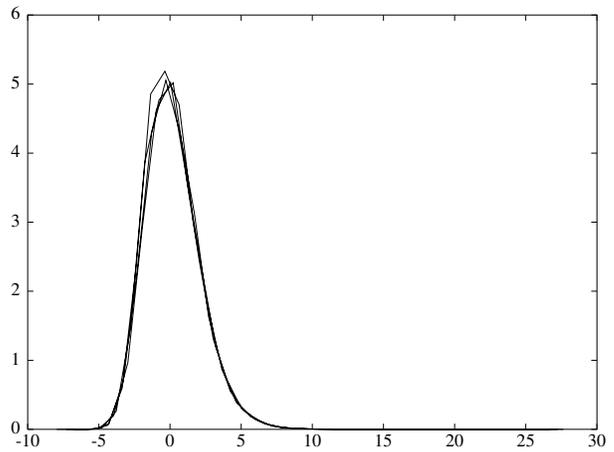,width=.7 \hsize,angle=-90}}
\caption{The curves of Fig. \ref{fig.P2} collapsed according to Eq. 
(\ref{I2}).}
\label{fig.coll}
\end{figure}

In analogy to \cite{abcpv01}, the statistical invariant $I^{(2)}$ for 
the passive scalar field is 
\begin{equation}
I^{(2)}(t) = \sum_n Z_n^{(2)} \langle |\psi_n (t)|^2 \rangle
\label{I2}
\end{equation}
where $Z_n^{(2)}$ is a left-eigenfunctions of the operator 
$\mathcal{P}^{(2)}$, with the scaling of the second order structure 
function of the forced problem. The invariance of $I^{(2)}$ is most 
easily demonstrated by re-scaling the decaying second order objects 
according to Eq. (\ref{I2}). The curves indeed collapse 
if the ordinate is shifted with the appropriate time dependence. This is shown
in Fig. \ref{fig.coll} and is analogous to Fig. 4 in \cite{cgp02}.
Thus in the language of the passive magneto-hydrodynamic model Eq. 
(\ref{mhdsabraa}), we have that 
\begin{equation}
\sum_n \left\langle |a_n|^2\right\rangle_f/k_n^\alpha
\left\langle|a_n|^2(t)\right\rangle/k_n^\alpha
\label{mhdI2}
\end{equation}
is a statistical integral of motion. 
To claim that the scaling of the magnetic field structure function 
$\langle |b_n|^2\rangle$ is a sub-leading zero mode of $\mathcal{P}^{(2)}$
is equivalent to claiming that
\begin{equation}
\sum_n \left\langle |b_n|^2\right\rangle/k_n^\alpha
\left\langle |a_n|^2(t)\right\rangle/k_n^\alpha
\label{mhdnoI2}
\end{equation}
is a statistical integral of motion. Clearly, in view of Fig. \ref{fig.coll} 
this cannot be the case and the collapse will not occur should the scaling 
exponent of $Z^{(2)}$ be replaced by another one
\footnote{Our claim is that there is only one unique scaling exponent $\beta_2$
(which is the scaling exponent of $Z^{(2)}$) which will make 
the quantity $\sum_n k_n^{\beta_2}/k_n^\alpha
\left\langle |a_n|^2(t)\right\rangle/k_n^\alpha $ a statistical integral of
motion. This can be proved if the velocity driving the passive field
is Kraichnan (i.~e. Gaussian delta-correlated in time), but for a generic
velocity field we know no way to  
rigorously prove this. One can however argue that the statistical integral
of motion should depend on the geometry of the correlation function and not 
on the details of the statistics of the advecting field.}.
In this line of thought, 
it is perhaps worthwhile pointing out that the collapse as seen in 
Fig. \ref{fig.coll} would not be possible should there be zero-modes with 
distinct scaling exponents. Indeed, as seen from Eq. (\ref{I2}), the collapse 
occurs provided $\langle |\psi_n(t)|^2\rangle$ ``falls'' precisely on the 
right eigenmode with a scaling identical to $Z_n^{(2)}$.

To conclude we emphasize that the linearity of the passive advection models 
assigns them to a narrow class of equivalence. The 
passive magneto-hydrodynamic model is in fact equivalent to a scalar advection 
model for which the statistical invariants have already been investigated in 
some detail \cite{cgp02}. In magneto-hydrodynamics, active and passive fields
have different scaling properties. The arguments that were used in the 
framework of thermal convection to account for the anomalous scaling of the
active field in terms of a passive auxiliary one do not carry over to 
magneto-hydrodynamics. The claim that one can nevertheless account for the 
scaling of the (active) magnetic field by considering sub-leading 
zero-modes of the operators propagating the decaying correlation functions was
proven wrong. In view of the form of the second order propagator, it is clear
that there are no zero mode but the one whose scaling is that of the
passive field. 
%Finally let us note that our work applies of course only in the 
%context of $2$nd order propagator.  Higher order propagators indeed
%possess more than one zero mode.

\acknowledgments
The authors gratefully acknowledge discussions with A. Celani, U. Frisch
and R. Pandit. T.~G. also wishes to thank Y. Cohen, I. Procaccia and
A. Pumir. D.~M. thanks D. Vincenzi. This work was partially done while 
D.~M. was visiting the Observatoire de la C\^ote d'Azur in Nice. 
The support of CEFIPRA under project number 2404-2 is acknowledged.  
T.~G. acknowledges financial support from the European Union under contract 
numbers HPRN-CT-2000-00162 and HPRN-CT-2002-00300.  D.~M. also wishes to
thank CSIR India for financial support.

\end{document}